\documentclass[11pt]{article}
\usepackage{amsmath,amssymb,amsthm}
\usepackage{color,mathtools,enumitem,xspace,xcolor}
\usepackage{bbm}
\usepackage[margin=0.88in]{geometry}
\usepackage[]{hyperref}
\hypersetup{colorlinks=true,urlcolor=blue,linkcolor=blue,citecolor=blue}
\allowdisplaybreaks
\usepackage[backend=biber, style=alphabetic, maxbibnames=4, minbibnames=3, doi=true, url=false, isbn=false, eprint=false]{biblatex}
\newcommand{\OTOCtwo}{OTOC$^{(2)}$\xspace}
\newcommand{\OTOCone}{OTOC$^{(1)}$\xspace}
\newcommand{\Und}{\mathcal{U}_{n,d}}

\newcommand{\eps}{\epsilon}

\newcommand{\ket}[1]{| #1 \rangle}

\DeclareMathOperator{\Tr}{Tr}

\newcommand{\poly}{\mathrm{poly}}

\makeatletter
\renewcommand{\paragraph}{%
    \@startsection{paragraph}{4}%
    {\z@}{1.9ex \@plus 1.3ex \@minus .6ex}{-1em}%
    {\normalfont\normalsize\bfseries}%
}
\makeatother

\makeatletter
\renewcommand\@fnsymbol[1]{\@arabic{#1}} 
\makeatother

\title{\bfseries \vspace{-6ex}A simplified version of the quantum \OTOCtwo problem\vspace{-1ex}}

\author{Robbie King$^{1,*}$, Robin Kothari$^{1,*}$, Ryan Babbush$^{1}$, Sergio Boixo$^{1}$,\\ Kostyantyn Kechedzhi$^{1}$, Thomas E. O'Brien$^{1}$, and Vadim Smelyanskiy\footnote{Google Quantum AI \hspace{15em} *Corresponding authors}}

\date{\vspace{-5.5ex}}

\begin{document}

\maketitle
\renewcommand{\abstractname}{\vspace{-5ex}}
\begin{abstract}
This note presents a simplified version of the \OTOCtwo problem that was recently experimentally implemented by Google Quantum AI and collaborators~\cite{Google2025}. We present a formulation of the problem for growing input size and hope this spurs further theoretical work on the problem.
\end{abstract}

\paragraph{Background and motivation.} In 2019, Google Quantum AI and collaborators~\cite{Google2019} experimentally implemented the Random Circuit Sampling problem for an input size that was beyond the ability of classical computers to simulate at the time. Later experiments have implemented sampling problems for input sizes believed to be classically intractable.

The primary drawback of a demonstration of quantum advantage using sampling problems is that its output is not efficiently verifiable. Two independent runs of a sampling task on an ideal quantum computer will almost certainly produce different answers. Verifiability is desirable because on the one hand it proves the output of the computation is correct, and on the other hand it is an essential characteristic of most practical applications. While the gold standard of verifiability is classical verifiability, where the quantum computer also outputs an efficiently checkable classical proof of the correct answer (as in the factoring problem), such problems remain beyond the reach of today's quantum computers. 
 
This motivates looking at problems that have a correct answer for each input, i.e., function computation problems, such as computing the expectation value of an observable on a quantum state. 
While this is weaker than classical verification, this opens up the possibility of verification by other quantum computers, including future quantum computers with lower error rates, and verification against nature itself in the context of quantum simulation. Useful deployments of quantum simulation, a flagship application of quantum computers, will be quantumly verifiable in the manner described above, but may not be classically verifiable.

Finally, to have experimental beyond-classical demonstrations, we need an algorithm along with specific hard instances. Worst-case hardness, such as BQP-completeness, is insufficient by itself since it does not guarantee that an average instance of the problem remains classically hard.
The recent experiment by Google Quantum AI and collaborators~\cite{Google2025} presents a problem satisfying these requirements (an expectation value problem that appears to have average-case quantum advantage), and we now describe a simplified version of this problem.

\paragraph{Problem definition.} 
Similar to Random Circuit Sampling, we start by picking a random quantum circuit from some ensemble. The details of the ensemble are not too important, but for concreteness, let us consider a random quantum circuit on a $2$D grid of $n = \ell \times \ell$ qubits. We sample Haar-random 2-qubit gates laid out in a brickwork pattern of 4 layers, alternating horizontal and vertical layers, where the first horizontal layer applies a 2-qubit gate between the qubits $(1,1)$ and $(1,2)$, and so on, whereas the second horizontal layer applies gates between $(1,2)$ and $(1,3)$ and so on. We call this distribution over circuits $\Und$, where $n$ is the number of qubits and $d$ is the depth of the circuit.

A typical task in quantum simulations of physical phenomena is to estimate (to some additive error $\eps=1/\poly(n)$) the expectation of some simple single-qubit observable $A$ on the state $U\ket{0^n}$. 
This is equivalent to measuring $\langle 0^n|U^\dagger A U|0^n\rangle$.
More generally, one could consider the well-studied \emph{time-ordered correlator} $\langle0^n|U^\dagger B U M|0^n\rangle$, where $B$ (called the ``butterfly'' operator) could be the Pauli $X$ operator on the qubit at location $(\ell,\ell)$, and  $M$ (the ``measurement'' operator) could be the Pauli $Z$ operator on the qubit at location $(1,1)$.
The issue is that for a random circuit $U \sim \Und$, these expectation values are extremely close to $0$ with high probability. This means there is no signal for the quantum computer to measure, and the trivial classical algorithm that just outputs $0$ for every input $U$ is almost always correct.

To have some non-zero expectation value, consider the second moment of the correlation operator $\langle 0^n|C^2|0^n\rangle$ where $C = U^\dagger B U M$. As before, let's say $U \sim \Und$, $B$ is Pauli $X$ on qubit $(\ell,\ell)$, and $M$ is Pauli $Z$ on qubit $(1,1)$. Let \OTOCone be the problem of estimating the expectation value $\langle 0^n|C^2|0^n\rangle$, where OTOC stands for \emph{Out-of-Time-Order Correlator}. If the depth of $U$ is too low and the light-cone of $U^\dagger B U$ does not reach $M$, then $U^\dagger B U$ and $M$ will commute, which ensures that $C^2 = \mathbbm{1}$ and $\langle 0^n|C^2|0^n\rangle = 1$. When the depth is large and the operators do not commute, then $C^2$ looks scrambled (i.e., behaves like a random unitary) and $\langle 0^n|C^2|0^n\rangle$ will be close to $0$. In the transition between these two regimes, it is conjectured that the OTOC exhibits inverse polynomial instance-to-instance fluctuations; see \cite[Sec. I]{Google2025}. However, it appears that in some cases, classical numerical methods are able to approximate \OTOCone efficiently.

This finally brings us to \OTOCtwo. We now consider the task of estimating $\langle 0^n|C^4|0^n\rangle$. On a quantum computer, we can use the fact that $M|0^n\rangle = |0^n\rangle$ to estimate \OTOCtwo by measuring the first qubit of the state $|\psi\rangle = C^2|0^n\rangle = (U^\dagger BU)M(U^\dagger BU)\ket{0^n}$. As before, if the depth of $U$ is low then $C^4 = \mathbbm{1}$ and so $\langle 0^n|C^4|0^n\rangle=1$. At intermediate circuit depth, $\langle C^4 \rangle$ takes instance-specific values that exhibit deviations from the Haar-random value.

More formally, in the \OTOCtwo problem for a distribution $\Und$, we are given a random $n$-qubit unitary $U$ drawn from $\Und$ and an $\eps>0$, and the goal is to output $\langle 0^n|C^4|0^n\rangle$ to additive error $\eps$. A quantum algorithm can solve this problem with gate complexity $O(nd/\eps^2)$, where $O(nd)$ is the gate complexity of implementing $U$, and with $O(1/\eps^2)$ repetitions we can estimate the expectation value to additive error $\eps$. (If deeper circuits are available this can be improved to $O(1/\eps)$ using amplitude estimation.) Ref.~\cite{Google2025} provides evidence that this problem is classically hard for the parameters chosen in the experiment. We conjecture that the general problem described above is classically hard for large $n$, i.e., there does not exist a classical algorithm with complexity $\mathrm{poly}(n,d,1/\eps)$. More specifically, for a 2D grid with $n=\ell \times \ell$, the problem should be hard for some $d \in \Theta(\ell)$ and $\eps=1/\mathrm{poly}(n)$.

More generally, one could consider the problem of approximating even higher moments of the correlation operator $\langle 0^n|C^{2k}|0^n\rangle$, 
which can be estimated on a quantum computer by measuring the first qubit of the state $|\psi\rangle = C^k|0^n\rangle$. One can also consider the problem of approximating the expectation value of this operator on the maximally mixed state, $\Tr(C^{2k})/2^n$.
In \cite[Sec. II]{Google2025}, arguments are provided about why certain classical algorithms based on Monte Carlo appear to encounter ``sign problems'' that become progressively worse as $k$ increases. This may provide some intuition about why these problems are hard for this class of algorithms.

\paragraph{Comparison with the experiment.} The problem presented above simplifies and abstracts away many of the hardware-specific choices of the experiment to make it easier to study theoretically. We now describe the differences between the version above and the experiment performed in the classically challenging regime: The actual experiment uses $n=65$ qubits that are not laid out in a perfect square grid as shown in \cite[Fig.~4]{Google2025}, and uses $d=23$ layers of 2-qubit gates. The experiment also uses a Pauli $Z$ operator as $M$, but uses a 3-qubit Pauli $X$ as $B$. The distribution over circuits used in the experiment is not Haar-random, but involves fixed 2-qubit gates (``iSWAP-like gates'') and random single-qubit gates from a specific distribution; this is described in \cite[Fig.~1]{Google2025}. The quantity reported in the experiment is not $\langle 0^n|C^4|0^n\rangle$, but a harder quantity where the easier-to-compute part of this quantity, called $\mathcal{C}^{(4)}_{\mathrm{diag}}$ in the paper, is subtracted off. The error metric used in the paper is not additive error; a signal-to-noise ratio is computed (equivalent to a correlation measure called Pearson correlation) between the ideal and experimental data sets. This does not precisely translate to a uniform additive error bound per instance, but we suspect that $\eps=0.001$ would pose a significant challenge for classical algorithms.

\paragraph{Acknowledgments.} We thank David Gosset, Jeongwan Haah, Tony Metger, and Rolando Somma for helpful discussions and comments on this note.

\printbibliography

\end{document}